\newcommand*{\wn}{cm$^{-1}$}
\newcommand{\etal}{\emph{et al.}}

\newcommand{\EFstate}{$EF^{1}\Sigma_{g}^{+}$}
\documentclass[aps,pra,twocolumn,superscriptaddress,10pt,floatfix]{revtex4-1}
\usepackage{longtable}
\usepackage{rotating}
\usepackage{graphicx}
\usepackage{amsmath}
\begin{document}

\title{Precision spectroscopy of high rotational states in H$_{2}$ investigated by Doppler-free two-photon laser spectroscopy in the $EF^1\Sigma_g^+$-$ X^1\Sigma_g^+$ system}
\author{G.D. Dickenson}
\affiliation{Department of Physics and Astronomy, LaserLaB, VU University, de Boelelaan 1081, 1081HV, Amsterdam, The Netherlands}
\author{E.J. Salumbides}
\affiliation{Department of Physics and Astronomy, LaserLaB, VU University, de Boelelaan 1081, 1081HV, Amsterdam, The Netherlands}
\affiliation{Department of Physics, University of San Carlos, Cebu City 6000, The Philippines}
\author{M. Niu}
\affiliation{Department of Physics and Astronomy, LaserLaB, VU University, de Boelelaan 1081, 1081HV, Amsterdam, The Netherlands}
\author{Ch. Jungen}
\affiliation{Laboratoire Aim\'{e} Cotton du CNRS, Universit\'{e} de Paris-Sud, 91405 Orsay, France}
\author{S.C. Ross}
\affiliation{Department of Physics and Centre for Laser, Atomic and Molecular Sciences, University of New Brunswick, P. O. Box 4400 Fredericton NB, Canada E3B 5A3}
\affiliation{Infrared Free Electron Laser Research Center, Tokyo University of Science, 2641 Yamazaki, Noda, Chiba 278-8510, Japan}
\author{W. Ubachs}
\email[Corresponding Author: ]{w.m.g.ubachs@vu.nl}
\affiliation{Department of Physics and Astronomy, LaserLaB, VU University, de Boelelaan 1081, 1081HV, Amsterdam, The Netherlands}

\begin{abstract}
Recently a high precision spectroscopic investigation of the $EF^{1}\Sigma^{+}_{g}$-$X^{1}\Sigma^{+}_{g}$  system of molecular hydrogen was reported yielding information on QED and relativistic effects in a sequence of rotational quantum states in the $X^{1}\Sigma^{+}_{g}$ ground state of the H$_2$ molecule [E.J. Salumbides \textit{et al.}, Phys. Rev. Lett. \textbf{107}, 043005 (2011)]. The present paper presents a more detailed description of the methods and results. Furthermore, the paper serves as a stepping stone towards a continuation of the previous study by extending the known level structure of the $EF^{1}\Sigma^{+}_{g}$ state to highly excited rovibrational levels through Doppler-free two photon spectroscopy. Based on combination differences between vibrational levels in the ground state, and between three rotational branches ($O$, $Q$ and $S$ branches) assignments of excited $EF^{1}\Sigma^{+}_{g}$ levels, involving high vibrational and rotational quantum numbers, can be unambiguously made. For the higher $EF^{1}\Sigma^{+}_{g}$ levels, where no combination differences are available, calculations were performed using the multi-channel quantum defect method, for a broad class of vibrational and rotational levels up to $J=19$. These predictions were used for assigning high-$J$ $EF$-levels and are found to be accurate within 5 \wn .
\end{abstract}

\maketitle

\section{Introduction}
\label{Sec:Introduction}
The hydrogen molecule and its deuterated isotopomers is the benchmark system for testing quantum \emph{ab initio}  calculations of molecular structure at ever increasing accuracy. While the first calculation of the dissociation energy of H$_{2}$ by Heitler and London~\cite{Heitler1927}, as the first application of quantum mechanics in molecular physics, was off by some 30\% from the experimental value, the accuracy has improved by many orders of magnitude since then. The calculations of Wolniewicz in the 1990s obtained an accuracy of 10$^{-7}$ for the binding energies of rovibrational levels in the H$_2$ ground state and long stood as a benchmark result~\cite{Wolniewicz1993,Wolniewicz1995}. In the past few years improved quantum \emph{ab initio} calculations of the Born-Oppenheimer potential and of adiabatic and non-adiabatic corrections for rovibrational levels of the $X^{1}\Sigma^{+}_{g}$ ground state of  H$_2$ have become available~\cite{Pachucki2009}. These were extended with detailed calculations of quantum electro-dynamical (QED) and relativistic effects to yield a theoretical value for the dissociation limit~\cite{Piszczatowski2009}, and for binding energies of all rovibrational levels in the $X^{1}\Sigma_{g}^{+}$ ground state of the H$_2$ molecule~\cite{Komasa2011} with an accuracy at the $10^{-8}$ scale.

These highly accurate level calculations were tested, in fact prior to the publication of the theoretical studies, by accurate measurements of the ionization potential (IP) of H$_2$~\cite{Liu2009} and D$_2$~\cite{Liu2010}. A measurement of the IP of HD followed~\cite{Sprecher2010} so that there are now tests for the advanced QED theory of the hydrogen molecule for all three stable isotopomers~\cite{Sprecher2011}. Previously, tests of QED in molecules had been restricted to one-electron systems like the HD$^+$ molecular ion~\cite{Koelemeij2007}. Note that in the IP measurements the absolute binding energy of the lowest ($J=0$) quantum state of parahydrogen is probed with respect to the energy of the ion. The contribution of QED and relativistic effects (hereinafter jointly referred to as QED effects) to the binding energy of the lowest rotational level in the specific case of H$_2$ is 0.7282 (10) \wn \cite{Komasa2011}; the accuracy of this calculation is implicitly tested in the IP measurements.

The availability of highly accurate QED calculations formed the motivation for  testing these phenomena in a sequence of rotational states in H$_2$, thereby seeking to reach high rotational quantum numbers. A prediction was made that the QED contributions to the binding energy should depend on $v$ and $J$ quantum numbers in an experimentally detectable amount~\cite{Komasa2011}. As for a possible detection strategy, the level structure of the H$_2$ $X^{1}\Sigma^{+}_{g}$ ground state can be probed directly through its purely rotational spectrum~\cite{Jennings1984} and through its vibrational spectrum~\cite{Bragg1982}, but due to the weak quadrupole nature of the transitions such measurements have not been performed in molecular beams with Doppler-free techniques. Recent Doppler-limited investigations using cavity enhanced techniques reached a relative accuracy of 10$^{-9}$, the most accurate determination of the quadrupole transitions to date~\cite{Cheng2012,Campargue2012}. These studies are limited to rotational levels $J \leq 5$ and pressure shifts need to be corrected for.

\begin{figure}
\includegraphics[width=1\linewidth]{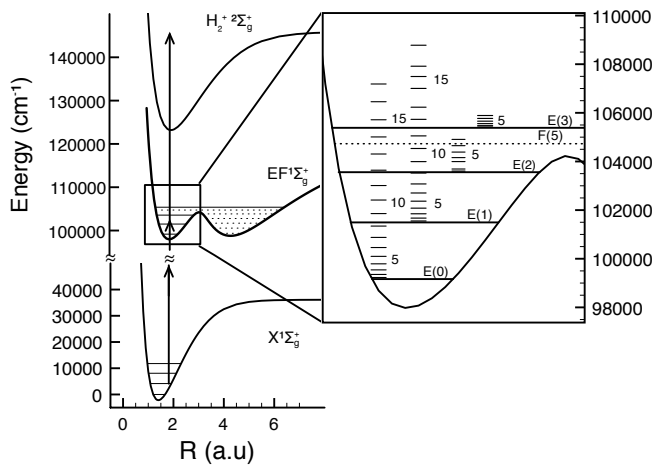}
\caption{Potential energy curves of the $X^{1}\Sigma^{+}_{g}$ ground state and the $EF^{1}\Sigma^{+}_{g}$ state of H$_2$, as well as the the $X^{2}\Sigma^{+}_{g}$ ground state of the H$_2^+$ ion. The excitation scheme is indicated with arrows. The first four vibrations of the inner well are indicated with solid lines, while the first six vibrations of the outer well are indicated with dashed lines. The rotational structure of the inner well vibrations is shown in the inset where the $J$ quantum number is indicated.}
\label{Fig:ExcitationScheme}
\end{figure}

In the present study we adopted the strategy of probing the ground state level structure via electronic transitions. In view of its lifetime in excess of 100 ns~\cite{Chandler1986} the $EF^{1}\Sigma^{+}_{g}$ state is a logical target and sensitive, high precision, Doppler-free spectroscopy of the $EF^{1}\Sigma^{+}_{g}$-$X^{1}\Sigma^{+}_{g}$ system has been amply demonstrated over the years~\cite{Eyler1987,Gilligan1992,Zhang2004,Yiannopoulou2006,Hannemann2006}. This scheme involving two-photon excitation from the $X^{1}\Sigma^{+}_{g}$ ground state is illustrated in Fig.~\ref{Fig:ExcitationScheme} and allows us to determine accurate level energies, as well as spacings between rovibrational levels, in the ground state of the H$_2$ molecule, up to high rotational quantum states. Initial results of this work have been published in a recent Letter~\cite{Salumbides2011}.

Using the present two-photon scheme, a limiting factor for further tests of binding energies in the $X^{1}\Sigma^{+}_{g}$ ground state is the unknown level structure of the $EF^{1}\Sigma^{+}_{g}$ state at high rotational quanta, $J \geq 5$. Transitions in the $EF-X$ system can be assigned as long as the level energies in the $EF^{1}\Sigma^{+}_{g}$ state are known, either from theory or from experiment. Yu and Dressler~\cite{Yu1994} have determined level energies for $EF(v, J)$ levels from coupled-channel calculations involving a Born-Oppenheimer potential for the $EF$-state~\cite{Wolniewicz1985}, adiabatic and non-adiabatic corrections~\cite{Wolniewicz1994}, as well as relativistic corrections~\cite{Wolniewicz1998}. These level energy calculations are limited to $J\leq 5$ and disagree with present measurements by a few 0.1 \wn\ for the lowest vibrational levels, and increase for higher vibrational levels. Up to $EF(15)$ (at 110 000 \wn\ ) the deviations between observed and calculated levels remain below 2.5 \wn. 

As part of the present study calculations of level energies were performed based on the multichannel quantum defect (MQDT) method~\cite{Ross1994}. Previously results were reported on $EF(v, J)$ level energies for $J\leq 5$ and the calculations are now extended to $J=19$. The results are used to assign the $EF$ levels, in particular, the high-$J$ levels in the various $EF-X$ bands observed.

On the experimental side a comprehensive high-resolution Fourier-transform spectroscopic study was performed in the visible and near-infrared range accessing a broad range of rovibrational levels in the $EF^{1}\Sigma^{+}_{g}$ state~\cite{Bailly2009} up to $v=28$. Of relevance for the present study are determinations of level energies up to $J=12$ for $E(v=0)$ and up to $J=5$ for $E(v=1-3)$; it should be noted that the absolute accuracies of level energies in Ref.~\cite{Bailly2009} depend on the determination of anchor lines in the $EF$ state by the two-photon laser experiments~\cite{Hannemann2006,Salumbides2008}. Those levels predominantly localised in the inner well are labelled as $E(v)$, while those localised in the outer well are labelled as $F(v)$, where $v$ enumerates the levels in each well, beginning at $v = 0$. Above the potential energy barrier between the two wells of the $EF$ state such labelling loses meaning and levels are referred to as $EF(v)$ where, for each $J$, $v$ enumerates all levels in energy order, beginning at $v = 0$.

Calculations of QED effects in the ground state can also be tested by measuring combination differences, i.e., rotational energy splittings of $\Delta J=2$ in the $X^{1}\Sigma^{+}_{g}$ ground state from pairs of $O(J+2)$ and $Q(J)$, and pairs of $Q(J)$ and $S(J-2)$ branch transitions. Still, such a comparison requires an unambiguous assignment of two-photon transitions in the $EF-X$ system involving states with high rotational quantum numbers. Therefore the present study targets two scientific issues: (i) to test QED calculations of $X^{1}\Sigma^{+}_{g} (v=0,J)$ ground state rovibrational levels, and (ii) to provide unambiguous assignments for transitions in the $EF-X$ system involving high rotational quantum numbers for $v > 0$.

\begin{figure}[b]
\includegraphics[width=1\linewidth]{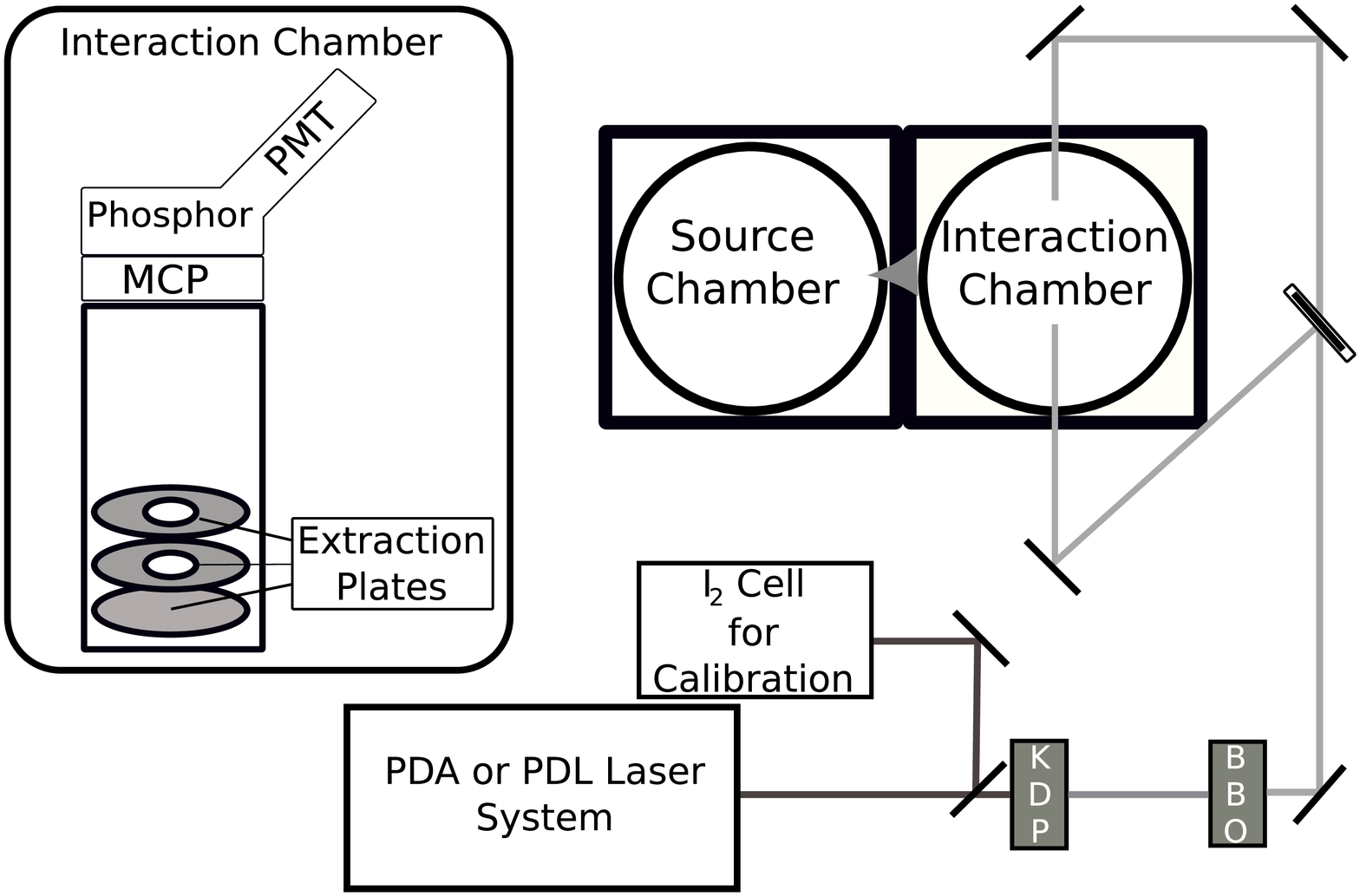}
\caption{Schematic layout of the experimental arrangement used in the experiments.}
\label{Fig:Setup}
\end{figure}

\section{Experimental details}
\label{Sec:ExperimentalSetup}

A schematic representation of the general experimental arrangement is shown in Fig.~\ref{Fig:Setup}.
%\subsection{High resolution laser system}
%\label{SubSec:PDA}
The precision metrology experiments were performed with a narrow-bandwidth pulsed dye amplifier (PDA) laser system.  It consists of a three-stage traveling-wave optical amplifier, pumped by a $Q$-switched Nd:YAG (yttrium aluminium garnet ) laser, and is seeded by the output of a continuous wave (cw) ring dye laser. The system delivers nearly Fourier-transform-limited pulses of 50 mJ and 5 ns duration, at a bandwidth of $\sim 100$ MHz. The instrument is tunable in the range 610-660 nm, while running on [2-[2-[4-(dimethylamino)phenyl]ethenyl]-6-methyl-4H- pyran-4-ylidene]-propanedinitrile (DCM) dye, although continuous scanning can only be accomplished over 1 cm$^{-1}$ intervals. A detailed description of this PDA system and its application in molecular spectroscopy is given in Ref.~\cite{Ubachs1997}.
%\subsection{Survey laser system}
%\label{SubSec:PDL}
A second laser used is a pulsed dye laser (Quanta Ray PDL-3), also pumped by a $Q$-switched Nd:YAG laser. This laser features a grating-based oscillator delivering a bandwidth of $\sim$0.06 \wn\ or $\sim$2 GHz. The continuous tunability over a broad frequency span makes this system suited for recording survey spectra. The laser was run on DCM laser dye and was tuned across the entire bandwidth of the dye with the aim of recording all detectable $EF-X$ transitions in this range.

%\subsection{Frequency up-conversion}
The pulsed output of both PDA and PDL lasers is up-converted in two stages to deliver pulsed radiation in the 204-220 nm range. Frequency doubling is accomplished in a KDP (potassium dihydrogen phosphate) crystal, and subsequent mixing in a BBO ($\beta$-barium-borate) crystal results in the third harmonic of the visible radiation. Typical output powers produced in the deep-UV are 0.4 mJ/pulse.

%\subsection{Calibration}
Frequency calibration of the lasers and determination of the transition frequencies is achieved by simultaneous recording of reference spectra of molecular iodine. For the PDL system a sufficient number of lines in a linear (Doppler-broadened) I$_{2}$ absorption spectrum are recorded in each scan to linearize the frequency scale and derive an absolute calibration~\cite{Gerstenkorn1978}. For precision measurement with the PDA system saturated I$_{2}$ absorption spectra are recorded using the cw seed-laser; the absolute frequencies of the hyperfine components are calibrated to 1 MHz accuracy~\cite{Xu2000,Iospec}. A linearized frequency scale in the PDA-precision measurements is produced from the transmission fringes of an etalon (free spectral range = 148.96 MHz), which is length stabilized by locking to a frequency stabilized HeNe laser.

%\subsection{Two-Photon Doppler Free}
Two-photon Doppler-free measurements of the $EF-X$ transitions are recorded using two counter-propagating deep-UV laser beams, both in the case of the precision measurements and for the survey spectra with the PDL-based laser. The UV-beam is split in two and arranged co-linearly as illustrated in Fig.~\ref{Fig:Setup}; exact counter-propagation is achieved by aligning the beams as part of a Sagnac interferometer~\cite{Hannemann2007}, to avoid Doppler-shifts. The high resolution PDA source is mildly focused with a single 1 m lens to produce sufficient intensity to excite the $EF-X$ transitions, but at the same time avoiding ac-Stark effects (see Sec.~\ref{SubSec:HighResolutionResults} for further details). For the PDL system a dual lens setup consisting of two 25 cm lenses is used for increased intensity at the interaction point to excite weaker transitions.
%\subsection{Ion Collection}

H$_2^+$ ions produced via 2+1 resonantly enhanced multi-photon ionization (REMPI) are detected and recorded as the laser is tuned. For the precision measurements an auxiliary 355 nm laser is used to ionize in a 2+1' REMPI scheme, \emph{i.e.} a two-color scheme. Pulse delay of this ionization laser by 30 ns helps to avoid ac-Stark effects. For the survey spectra obtained with the PDL system a one-color scheme is used. Ions are accelerated in a time-of-flight tube, 50 cm in length, towards a multichannel plate (MCP). The signal on the MCP is converted by a phosphor screen and a photo-multiplier tube, gated in a boxcar and digitally stored. The voltages on the extraction plates are pulsed to measure under field-free conditions, thereby avoiding dc-Stark effects.

%\subsection{The generation of rovibrationally excited H$_{2}$}
In order to produce high rotational angular momentum states (and vibrational excitation up to $v=3$) of the H$_2$ molecule a sequence of reactions is used, occurring in a beam of HBr molecules. A pulsed solenoid valve (General Valve, Series 9) in the source chamber releases a pulse of HBr molecules that enters through a (1.5 mm diameter) skimmer into a differentially pumped interaction chamber.
%The pressure in the source chamber, evacuated by a diffusion pump, rises from $8\times10^{-7}$ to $1\times10^{-4}$ %mbar when the pulsed valve is operational, while the interaction chamber evacuated by a turbo pump remains below %$10^{-6}$ mbar. Note that these pressures give no indication on the density in the molecular beam.
In a first photolysis reaction HBr molecules are photolyzed by the same UV laser beam (in the range 204-220 nm), which is also used as the spectroscopy laser:
\begin{equation}
\mathrm{HBr} + h\nu_{\mathrm{UV}} \rightarrow  \mathrm{H} + \mathrm{Br}
\end{equation}
resulting in H atoms with a high kinetic energy of up to 2 eV, which is then sufficient to overcome the reaction barrier~\cite{Aker1989} to undergo a secondary exothermic reaction:
\begin{equation}
\mathrm{HBr} + \mathrm{H} \rightarrow \mathrm{H}_{2} + \mathrm{Br}
\end{equation}
The reaction dynamics of such processes have been investigated in detail and it is well established that these result in product internal state distributions with rotationally and vibrationally excited H$_{2}$($v,J$) molecules~\cite{Rinnen1991,Huo1991,FernandezAlonso2000,Pomerantz2004}. Heck and co-workers~\cite{Heck1995} used a similar reaction dynamics scheme, using DI as a precursor gas, with the explicit goal to determine transition frequencies involving high-$J$ levels ($J=26$) in D$_2$; the latter study was not performed under Doppler-free conditions.

\begin{figure}[t]
\includegraphics[width=1\linewidth]{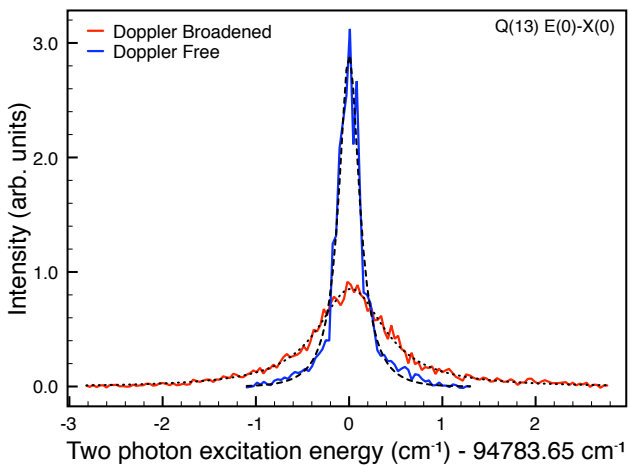}
\caption{(Color online) Measurement of the $Q(13)$ $EF^{1}\Sigma^{+}_{g}$-$X^{1}\Sigma^{+}_{g}$ (0,0) two-photon transition with the PDL system under Doppler-broadened and Doppler-free conditions. For details see text.}
\label{Fig:Q13}
\end{figure}

As an example, in Fig.~\ref{Fig:Q13} recordings with the PDL system of the $EF-X$ (0,0) $Q(13)$ transition are displayed. One of the measurements is recorded with a single laser beam yielding a Doppler-broadened profile ($\sim 0.9$ \wn\ width), while a second measurement was performed with two counter-propagating beams resulting in a Doppler-free spectrum; the recordings are area normalized to one. Both measurements were conducted under similar focussing conditions, at intensity levels where ac-Stark broadening is not significant for these linewidths. The width of the Doppler-free signal ($\sim 0.3$ \wn) is caused by the laser line-width, accounting for frequency tripling and two-photon excitation. Figure~\ref{Fig:Q13} shows the advantage of the Doppler-free geometry in the enhancement of signal strength. Due to the large collisional energy of the H atoms (2 eV), H$_{2}$ products are formed with a spread of translational energies resulting in the Doppler-broadened line shape.
%(\emph{Please discuss these issues with Edcel and Mingli and come with a quantitative model on the kinetics. By my calculations %the H atoms gain more than 2 eV of kinetic energy in the photolysis step. If we assume that the collision is completely elastic %and include the exothermicity of the reaction, the H$_2$ molecules formed have 1.93 eV . According to the reaction dynamics %studies of Aker \etal \cite{Aker1989} roughly half of this energy will go into rotation and vibration of the molecule. Hence %the H$_2$ molecules have 0.96 eV of rovibronic energy. This should be the peak of the energy distribution since we observe the %J=17 v=0 level which requires 1.84 eV of excitation energy.})

\section{MQDT calculations}
\label{Sec:MQDT}

The assignments of the $EF$ energy levels were made on the basis of first-principles non-adiabatic rovibronic MQDT calculations. These were carried out exactly as in a previous study~\cite{Ross1994}, by use of the same theoretical approach \cite{Ross1994a,Ross1994b} and the same input parameters for the computations.

Briefly,  in Ref. \cite{Ross1994} bond length- and energy-dependent quantum defect matrices {\boldmath $\mu$}$(E,R)$ were extracted from the then best available \textit{ab initio} clamped nuclei potential energy curves of Refs. \cite{Wolniewicz1985,Dressler1984,Kolos1982}. Three distinct sets of  matrices, corresponding to $^1\Sigma_g$ ($3\times 3$ matrix),  $^1\Pi_g$ ($2\times 2$ matrix), and  $^1\Delta_g$ ($1\times 1$ matrix) molecular symmetry, respectively, were used, each composed of singly excited $1\sigma_g\epsilon\ell\lambda$ and doubly excited $1\sigma_u\epsilon\ell'\lambda$ Rydberg channels. $1\sigma_g$ and $1\sigma_u$ denote the ground state and first excited state H$_2^+$ core orbital, and the partial wave indices $\ell\lambda$ and $\ell'\lambda'$ ($\ell\lambda,\ell'\lambda'\leq 2$) for the Rydberg electron are chosen such as to yield the required molecular symmetry. Each quantum defect matrix element is a smooth function of bond length and energy. The matrix elements were adjusted to the quantum-chemical potential energy curves in such a way that when re-injected into a clamped-nuclei MQDT calculation, they reproduced the potential energy curves as best as possible. 

Rovibrational motion and ro-vibronic interactions were introduced by means of the frame transformation described in Ref. \cite{Ross1994b}. This transformation converts the clamped-nuclei quantum defect matrices into much larger rovibronic quantum defect matrices which typically are of dimension $\approx 400 \times 400$  and which account for the electronic as well as for the rovibrational degrees of freedom, and which, in particular, also include non-adiabatic electron-core interactions. Quantum defect techniques described in Refs. \cite{Ross1994a,Ross1994b,Ross1994} yield the desired non-adiabatic level energies.   This purely \textit{ab initio} approach reproduced more than 270 excited-state singlet \textit{gerade} levels of H$_2$ in the range $0 \leq J \leq 5$ with an overall rms error of about 6 cm$^{-1}$ \cite{Ross1994}, just slightly larger than the rms error of about 4 cm$^{-1}$ obtained simultaneously by a coupled differential equations method for the same set of levels \cite{Yu1994}.  

\begin{figure*}[t]
\includegraphics[width=1\linewidth]{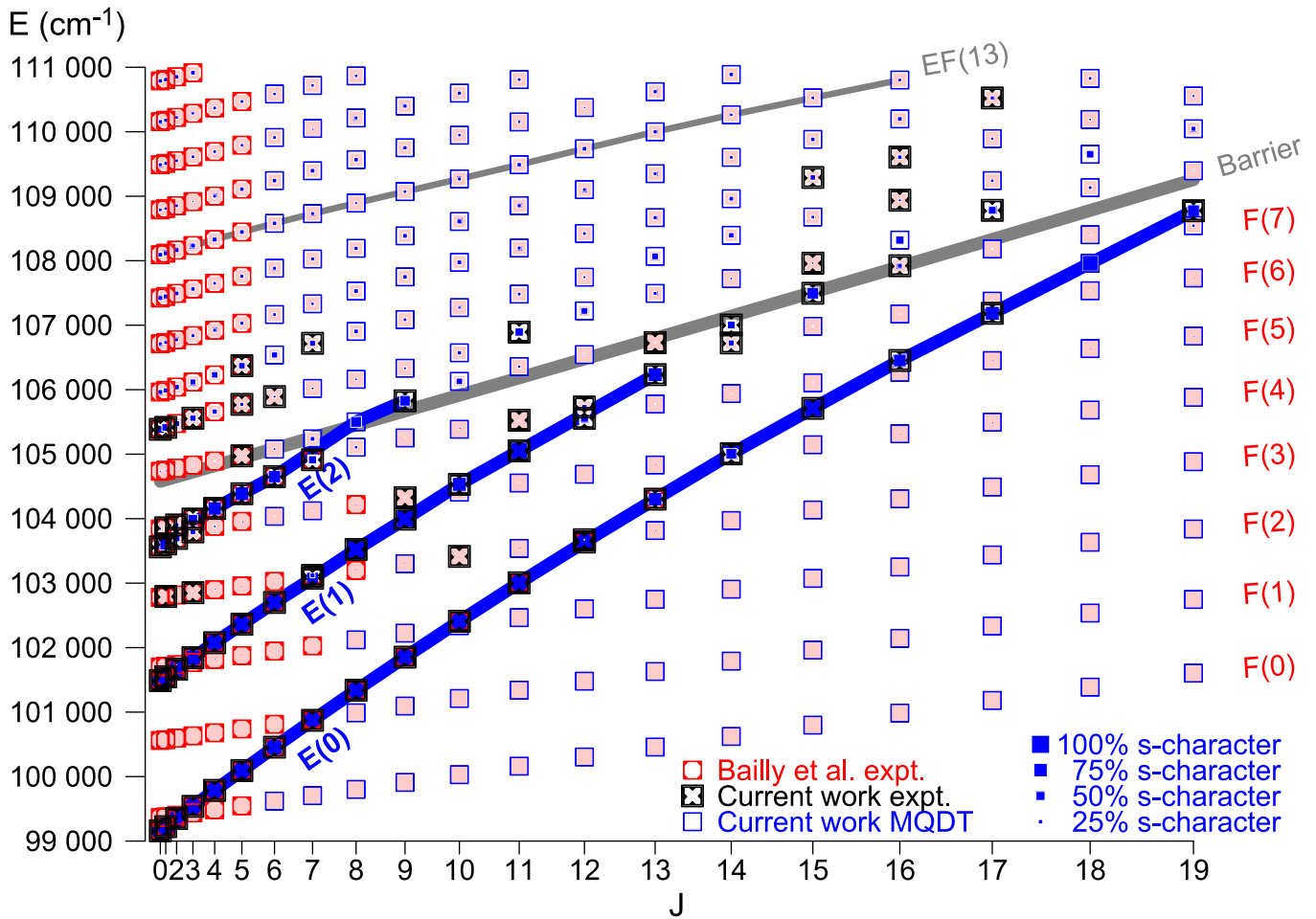}
\caption{(Colour online) Calculated energies and electronic character of the rovibronic energy levels of the $EF^{1}\Sigma_{g}^{+}$ of H$_{2}$ shown plotted vs. $J$. The type of square used to show a level indicates the level's status: Levels observed by Bailly \etal \cite{Bailly2009} are shown by squares with color-indented corners (red online), those observed in the present work are shown by double squares with indented sides (black online), while unobserved levels calculated in the present work are shown by simple squares (blue online). As can be seen some levels were observed in both Ref.~\cite{Bailly2009} and the current work. The calculated electronic character of each level is indicated by the size of the shaded square areas inside each outer square: Heavy shading (blue online) indicates the percentage of ($1\sigma_g$) $\epsilon s \sigma$ Rydberg $s$ $^{1}\Sigma_{g}^{+}$ channel, while light shading (pink online) indicates the amount of doubly excited ($1\sigma_u$) $\epsilon p \lambda$ (0 $\leq  \lambda \leq$ 2) channels as well as of singly excited ($1\sigma_g$) $\epsilon d \lambda$ channels. A fully filled square corresponds to 100\% character. To guide the eye, the rotational progressions $E(0)$, $E(1)$ and $E(2)$ associated with the inner potential well of the $EF$ state are connected by shaded lines. Heavy shading (grey online) indicates the potential energy barrier and the $EF(v=13)$ level which is above the barrier. Note that on the scale of the figure the largest observed-MQDT deviations are less than one third of the thickness of the lines used to draw the axes. See text for more details.}
\label{Fig:LevelEnergiesMQDT}
 \end{figure*}

This accuracy, which at the time corresponded to the state of the art of theory, is sufficient for the present purpose of assigning the new levels with $J > 5$, and therefore we have used the old approach without change. Improvements are possible but deferred at this time. They would include: (i) use of the improved first-principles potential energy curves available today; for instance, the potential energy curve for the $EF^{1}\Sigma_{g}^{+}$ state reported in Ref.  \cite{Wolniewicz1994} and not accessible to the authors of Ref. \cite{Ross1994} is lowered with respect to the curve used in \cite{Ross1994} by 1 up to 15 cm$^{-1}$, depending on the $R-$value; (ii) an improved fit of the quantum-chemical data by the clamped-nuclei quantum defect matrices; indeed, the matrices derived in Ref. \cite{Ross1994} reproduced the \textit{ab initio} potential energy curves only to within about 4 to 8 cm$^{-1}$; and (iii) inclusion of a larger number of channels beyond $\ell,\ell' \leq 2$ in the clamped-nuclei MQDT treatment.  

Figure~\ref{Fig:LevelEnergiesMQDT} shows all levels in the range 99\,000 to 110\,000 \wn\ for $0\leq J \leq19$. The corresponding numerical information, along with existing high-precision experimental data are presented in Table I of the supplemental material~\cite{SuppHBrPRA}. The figure displays the MQDT level energies and, as detailed in the caption, an indication as to whether the levels were seen in the present work, the work of Bailly \etal\ \cite{Bailly2009}, or both. Shading of the energy levels in the figure indicates the character of the MQDT wave function for each level, either $s$-character (blue shading online) or combined doubly-excited and $d$-character (light pink shading online). The potential energy barrier between the inner and outer wells, including the centrifugal $J(J+1)$ term, is shown by the essentially straight thick line (grey online) rising up from near 104\,500 \wn\ at $J=0$. In the region near and below this potential energy barrier those levels with predominantly $s$-character nicely line up to form the $E(v)$ inner well vibrational states. These levels are connected by wide shaded lines (light blue online) and are labelled $E(0)$, $E(1)$, and $E(2)$. Well below the potential energy barrier levels with predominantly doubly excited character line up to form the $F(v)$ outer well vibrational states, particularly at higher $J$ values. These are labelled $F(0-7)$. In the region well above the potential energy barrier the levels have predominantly doubly excited character and are now easily identifiable as vibrational states of the combined $EF$ well. One of these, $EF(13)$, is labelled and connected by a line (grey online). Beginning at $J=0$ in the region 105\,500-106\,000 \wn , just above the potential energy barrier, a renewed concentration of $s$-character occurs. For each $J$ this $s$-character is shared between two adjacent levels for low values of $J$. The lower of each of these pairs was identified as belonging to an "$E(3)$" level in Ref.~\cite{Bailly2009}, but since the $s$-character is well less than 50\% in each case the physical assignment no longer holds, although we retain the $E(3)$ in the nomenclature for the purpose of comparison with literature values. This concentration of $s$-character above the barrier is reminiscent of a shape resonance embedded in a dissociation continuum, except that in the present example the continuum is replaced by the relatively dense discrete manifold of levels associated with the outer $F$ state well. Another recurrence can be faintly seen at low $J$ around 107\,500-108\,000 \wn\ in Fig.~\ref{Fig:LevelEnergiesMQDT}. 

\section{Precision metrology}
\label{SubSec:HighResolutionResults}
\begin{figure}[t]
\includegraphics[width=1\linewidth]{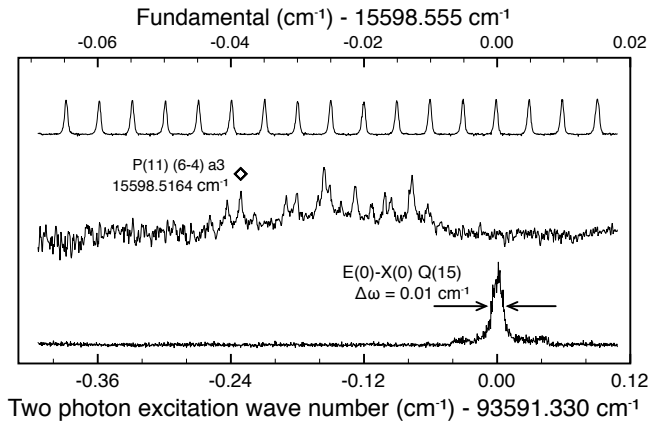}
\caption{The $EF^{1}\Sigma^{+}_{g}$-$X^{1}\Sigma^{+}_{g}$ (0,0) two photon $Q(15)$ transition recorded with the high-resolution PDA system. The fringes of a frequency-stabilized etalon with FSR = 148.96 MHz is used to linearise the scan while the known I$_{2}$ hyperfine component, marked with a $\diamond$, provides an absolute calibration. The transition frequency axis (lower axis) is exactly six-fold the fundamental frequency axis.}
\label{Fig:Q15}
\end{figure}

Precision metrology results on 21 transitions in the $EF^{1}\Sigma^{+}_{g}$-$X^{1}\Sigma^{+}_{g} (0,0)$ band were obtained via 2+1' REMPI from measurements on a line-by-line basis with the high resolution PDA system. As an example a recording of the $Q(15)$ two-photon line is shown in Fig.~\ref{Fig:Q15}, where simultaneous recordings of markers of the frequency-stabilized etalon and the I$_{2}$ saturated absorption spectrum provide the calibration. The spectral intensity of the light source is sufficient to detect $S(J)$ and $O(J)$ branch transitions, known to be considerably weaker than $Q(J)$-branch transitions~\cite{Marinero1983}. The $\Delta J=2$ ground state rotational level spacings were determined by taking combination differences between $Q(J)$ and $S(J-2)$, and $Q(J)$ and $O(J+2)$ transitions. 

The identification of the two-photon lines in the $EF-X$ (0,0) band derives from two consistent methods: (i) the derived $\Delta J=2$ combination differences in $X^{1}\Sigma^{+}_{g}, v=0$ match the calculations of Komasa~\etal~\cite{Komasa2011}; (ii) the observed transition frequencies match differences between the highly accurate (experimentally determined) $EF$-level energies by Bailly~\etal~\cite{Bailly2009} and the $X$-ground state level energies of Ref.~\cite{Komasa2011}. The $EF^{1}\Sigma^{+}_{g}, v=0, J=13$ level of Ref.~\cite{Bailly2009} is found to be inconsistent with observed combination differences, indicating a mis-assignment in Ref.~\cite{Bailly2009}. The $J=13$, $J=15$, and $J=16$ levels were unambiguously assigned and measured at high precision with the PDA system. In view of the limited tuning range of the PDA system the $J=14$ and $J=17$ levels were not found. A listing of transition frequencies, corrected for the ac-Stark effect, is given in Table II in the Supplemental Material~\cite{SuppHBrPRA}.

\begin{figure}[b]
\includegraphics[width=1\linewidth]{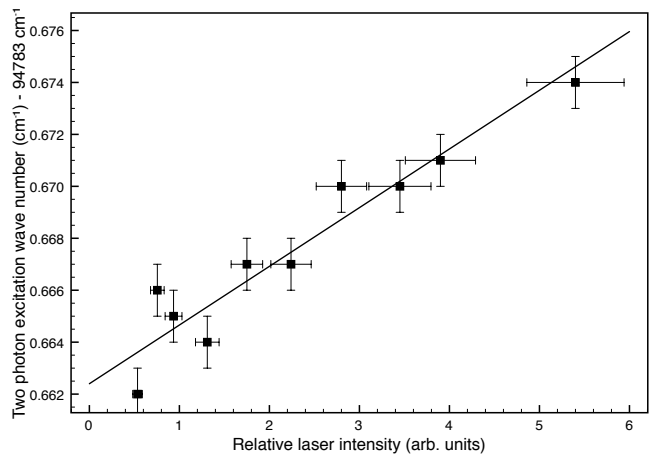}
\caption{The measured $Q(13)$ $EF^{1}\Sigma^{+}_{g}$-$X^{1}\Sigma^{+}_{g}$ (0,0) two photon transition frequency at several different input powers. An extrapolation of a linear fit yields the un-shifted transition frequency.}
\label{Fig:Q13StarkPlot}
\end{figure}

%\subsubsection{Uncertainty analysis}
%\label{SubSubSec:UncertaintyAnalysis}
Systematic investigations were pursued into the uncertainties of the transition frequencies measured with the PDA-system. These are summarised in Table~\ref{Tab:ErrorBudget}, and the most significant contributions are the chirp-induced frequency shifts in the PDA, the ac-Stark or power-induced effects, statistical fitting, absolute calibration of the frequency scale, and Doppler effects.

A major contribution to the measurement uncertainty derives from the frequency chirp due to time-dependent gain in the dye amplifiers. This phenomenon has been extensively characterized for the presently used PDA system~\cite{Eikema1997}. Based on those investigations we estimate an upper limit to a possible chirp-induced frequency shift. From measurements on the Lyman and Werner bands in H$_2$ it was found that at the edge of dye tuning curves this systematic effect can be pronounced~\cite{Bailly2009}. Hence we include a conservative limit of 150 MHz due to the chirp in the error budget.
A measurement of the $Q(5)$ $EF^{1}\Sigma^{+}_{g}$-$X^{1}\Sigma^{+}_{g}$ (0,0) transition was found to agree within 10 MHz with a previously reported measurement taken at higher accuracy, using a laser system with Fourier-transform laser pulses at longer duration~\cite{Salumbides2008}, indicating that the estimate of the chirp is conservative.

The ac-Stark effect was found to broaden, shift, and cause an asymmetry in the measured line shapes of the $EF-X$ two-photon transitions. This phenomenon was assessed by measuring each transition in the $EF^{1}\Sigma^{+}_{g}$-$X^{1}\Sigma^{+}_{g}$ (0,0) band as a function of intensity of the spectroscopy laser in the deep UV; as an example results for the frequency shift are shown in Fig.~\ref{Fig:Q13StarkPlot} for the $Q(13)$ transition. An extrapolation to zero input power through a weighted linear fit yields the un-shifted transition frequency. The $y$-axis error bars are the statistical uncertainty in each measurement point (averaged over three recordings) and the $x$-axis error bars represent the error in the average pulse energy measurement estimated at 10 \% of the measured value. The power measurement was done with a UV sensitive photodiode and provides a relative power scale. The uncertainty in the zero power extrapolation fell between 0.0005 and 0.0015 \wn. We take the upper limit, corresponding to 45 MHz, as an estimate of the uncertainty in the ac-Stark shift. 

The precision measurements are performed at low laser power (200-400 $\mu$J/pulse) inducing the two-photon excitation, and ionization by a 355 nm pulsed laser, delayed by 30 ns. In the case of the $Q(6)$ transition the ac-Stark effect was measured on an absolute power scale. It was found that for the input powers and focussing conditions of the PDA system (200-400 $\mu$J/pulse focussed with a 1 m lens), the ac-Stark shift amounts to approximately 150 MHz. The ac-Stark shift coefficient for the $Q(6)$ transition results in a value of 1 MHz/(MW/cm$^{2}$) which agrees with the more accurate result of Hannemann \etal~\cite{Hannemann2006}.

The excitation is performed under field-free conditions to prevent dc-Stark effects shifting the transition frequency. Pulsed voltages of 1465 V and 2000 V are applied to the extractor and repeller plates respectively in order to collect the ions produced from the 2+1 REMPI process. Nevertheless, stray fields may be present that can cause a shift in the transition frequency and a conservative estimate of 10 MHz is included in the error budget.

The absolute calibration, derived from a Doppler-free I$_{2}$ spectrum and a linearisation procedure based on the transmission markers of the stabilized etalon, is estimated at $\sim$5 MHz in the two-photon transition frequency. Doppler shifts, possibly caused by deviations from a perfect alignment of counter-propagating laser beams, are avoided in a Sagnac configuration~\cite{Hannemann2007}; residual angular mismatch amounts to some 1 MHz in the calibration uncertainty~\cite{Hannemann2006}. The profiles of the observed lines in the Doppler-free geometry are essentially Lorentzian and the observed widths (on average 300 MHz ) are determined by the harmonically converted line width of the PDA-system. Line fitting errors were assessed by fitting with both a Voigt and a Lorentzian function. Deviations between fits with the different functions were found to be at the 5 MHz level. The statistical uncertainty represents differences on a daily basis and was found to be at the 30 MHz level due to the laser linewidth and the signal-to-noise ratio.

In Table~\ref{Tab:ErrorBudget} a summary of the error budget is given resulting in a combined uncertainty of 0.005 \wn\ or 160 MHz by adding individual contributions in quadrature sum.

\begin{table}[t]
\caption{Error budget for the precision metrology measurements obtained with the high-resolution PDA system. See text for details.}
\label{Tab:ErrorBudget}
\begin{center}
\begin{tabular}{c|c|c}
\hline
\hline
	&Uncertainty MHz  & Uncertainty \wn  \\
\hline
  Chirp& 150 & 0.005 \\
  ac-Stark& 45 & 0.0015  \\
  Statistical& 30 & 0.001 \\
  dc-Stark& 10 & 0.0003 \\
  Line fitting & 5 & 0.0002 \\
  I$_{2}$ calibration & 5 & 0.0002 \\
  Etalon non-linearity& 5 & 0.0002 \\
  Residual Doppler& 1 & 0.00003 \\
  &  &  \\
Combined error& 160 & 0.005 \\
\hline
\hline
\end{tabular}
\end{center}
\end{table}

\section{Test of QED in the H$_2$ ground state}
\label{Sec:MeasurementCombinedQEDEffects}

%The results of the precision measurements are used to extract ground state
%X $^1\Sigma^+_g$ rotational level energies by subtracting the measured
%EF-X transition frequencies from the known level energies of the EF $^1\Sigma^+_g$  
%($v=0, J$) state from Ref.~\cite{Bailly2009}. The accuracy of the resulting ground state
%level energies enabled us to derive energy corrections beyond the nonrelativistic
%energy contributions. 
From a combination of the precision $EF-X$ transition frequencies and the known $EF$ level energies~\cite{Bailly2009} ground state $X^1\Sigma^+_g$ rotational level energies can be derived. The accuracy of these results enables us to derive corrections to the ground state rotational levels that go beyond the nonrelativistic energy contributions. Experimentally it is not possible to disentangle the various contributions to the rotational excitation energies. However, starting with the experimental $EF$ energies and subtracting the most accurate \emph{ab initio} nonrelativistic energies of the ground state, comprising of the Born-Oppenheimer, adiabatic, and nonadiabatic contributions~\cite{Pachucki2009}, QED and relativistic  corrections could be derived. This procedure yields the (differential) QED and relativistic corrections to the binding energies in the $X^1\Sigma^+_g$ ground state for the rotational sequence $J=2-16$ with respect to the $J=0$ corrections.

%Alternatively, the QED effects in the X$^{1}\Sigma^{+}_{g}, J$ ground state rotational manifold of H$_2$ can be differentially probed by using %experimentally determined combination differences between Q($J$) and S($J+2$) and Q($J$) and O($J-2$) branch transitions. By combining %transitions with a common excited state level it is possible to measure ground state rotational level spacings separated by $\Delta J$=2. Again by %subtracting the BO, adiabatic and non-adiabatic effects~\cite{Pachucki2009}, differential $J$-dependent experimental values for the QED effects in %the ground state result.

Alternatively, ground state rotational energy splittings can be derived by using experimentally determined combination differences between $EF-X$
$Q(J)$ and $S(J-2)$, and $Q(J)$ and $O(J+2)$ branch transitions. By combining transitions with a common excited state level it is possible to measure ground state rotational level spacings separated by $\Delta J=2$.  In similar fashion, by subtracting the BO, adiabatic, and nonadiabatic
effects~\cite{Pachucki2009}, we derive differential $J$-dependent experimental values for QED and relativistic effects in the ground state rotational splittings.

%There are several issues on these analyses of QED-effects that bear mentioning. The direct derivations of the QED-effects are dependent on the accuracy of the EF level energies determined in Ref.~\cite{Bailly2009}, which are in turn referenced to the two anchor lines as measured in the EF-X system~\cite{Hannemann2006}. The second method probing the QED effects from combination differences is independent of the accuracy of the EF level energies, although prior unambiguous assignments of the specific EF levels is a pre-requisite. Both methods addressing the QED-effects are referenced to the X$^{1}\Sigma^{+}_{g}, v=0, J=0$ level in the ground state. This level exhibits a global QED shift with respect to the dissociation energy of +0.7283 (10) \wn\ and this value is equally contained in the entire rotational manifold. The accuracy of this theoretical value has been tested in an entirely different experiment sensitive to the absolute binding energy of the X$^{1}\Sigma^{+}_{g}, v=0, J=0$ level~\cite{Liu2009}. The present work tests differential QED-effects in a sequence of rotational levels, up to $J=16$ in H$_2$.

There are several issues in these analyses of the QED and relativistic effects that bear mentioning. Although, the two methods are equivalent in principle, the first method mentioned is dependent on the accuracy of the $EF$ level energies determined in Ref.~\cite{Bailly2009}, which are in turn referenced to the two anchor lines as measured in the $EF-X$ system~\cite{Hannemann2006}. The second method probing the ground state rotational energy splitting is independent of the accuracy of the $EF$ level energies, although prior unambiguous assignments of the specific $EF$ levels is a pre-requisite. The experimentally derived QED and relativistic corrections obtained from both methods are referenced to the  $X^1\Sigma^+_g$, $v = 0, J = 0$ level in the ground state. This level exhibits a combined QED and relativistic shift of +0.7282 (10) \wn\ towards the dissociation limit \cite{Komasa2011}, and the entire rotational manifold is shifted by this amount in the absolute sense. The accuracy of this theoretical QED and relativistic contributions has been tested in an entirely different experiment sensitive to the absolute binding energy of the  $X^1\Sigma^+_g$ $v = 0, J = 0$ level~\cite{Liu2009}. The present work tests differential QED and relativistic effects in a sequence of rotational levels, up to $J = 16$ in H$_{2}$.

The experimentally derived QED and relativistic corrections are in excellent agreement with the recent
most accurate calculations of Komasa \etal~\cite{Komasa2011} which includes the lowest-order relativistic corrections
and QED corrections up to $\alpha^3$-order, where the accuracy of the calculation is limited by
the estimated higher-order $\alpha^4$ QED contribution.

\begin{figure}[t]
\includegraphics[width=1\linewidth]{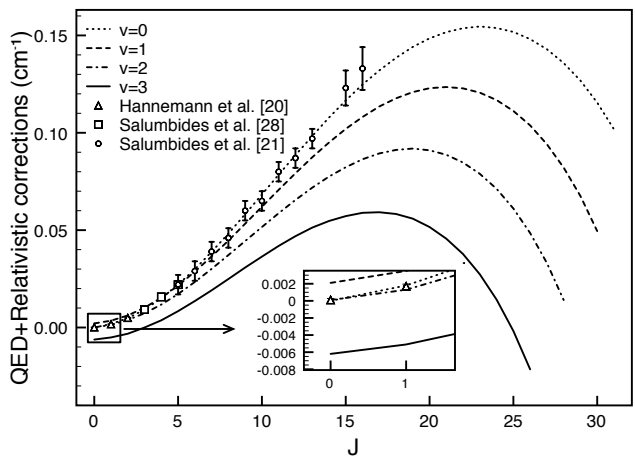}
\caption{A plot of the QED effects (including relativistic effects) in the $X^{1}\Sigma^{+}_{g}$ ground state of H$_2$ as a function of rotational quantum number $J$. Curves represent calculated differential QED effects (with respect to the $v=0, J=0$ level) for the four lowest $v=0-3$ levels as calculated by Komasa~\etal~\cite{Komasa2011}. The data points with plotted uncertainties represent the experimental data as described in the text.}
\label{Fig:QEDAndRelCorrections}
\end{figure}

%The results of the experimentally determined QED-effects in the H$_2$ ground state are presented in Fig.~\ref{Fig:QEDAndRelCorrections}. Also shown in the figure are the theoretical predictions for QED-effects in rotational sequences for vibrations $v=0-3$. For the $v=0$ sequence very good agreement is found between experimental and theoretical values; hence the data of Fig.~\ref{Fig:QEDAndRelCorrections} demonstrate the correctness of QED theory for the H$_2$ molecule. Future precision tests on rotational sequences for $v > 0$ can be performed, also using the EF-X system in H$_2$, if such quantum states can be sufficiently populated. A limiting factor for further precision measurements on QED-effects are the unknowns in the level structure of the EF state, which even makes the mere assignment of lines difficult. The recording of survey spectra as presented in the next section are meant to address this specific problem.

The results of the experimentally determined QED and relativistic corrections in the H$_{2}$ ground state are presented in Fig.~\ref{Fig:QEDAndRelCorrections} (experimental data is given in Table II of Ref.~\cite{Salumbides2011}). Also included are the theoretical predictions for QED and relativistic effects in rotational sequences for vibrations $v = 0 - 3$. At present the predictions~\cite{Komasa2011} and measurements~\cite{Salumbides2011} agree within the experimental uncertainty, including the results of Wolniewicz~\cite{Wolniewicz1995} which cover the limited range of quantum states below $J=10$. 
The $EF$ state can also be used for testing theory for rotational sequences of vibrationally excited levels of the electronic ground state, provided that $v > 0$ levels can be populated. A limiting factor for precision measurements towards further tests of QED and relativistic effects is the difficulty in the assignment of the transitions to the highly excited rotational quantum states ($J>10$) of the $EF$ state. The recording of survey spectra as presented in the next section is meant to address this specific problem.

\section{Survey spectra and further assignment of $EF$-levels}
\label{SubSec:ResultsSurveySpectrum}

\begin{figure}[b]
\includegraphics[width=1\linewidth]{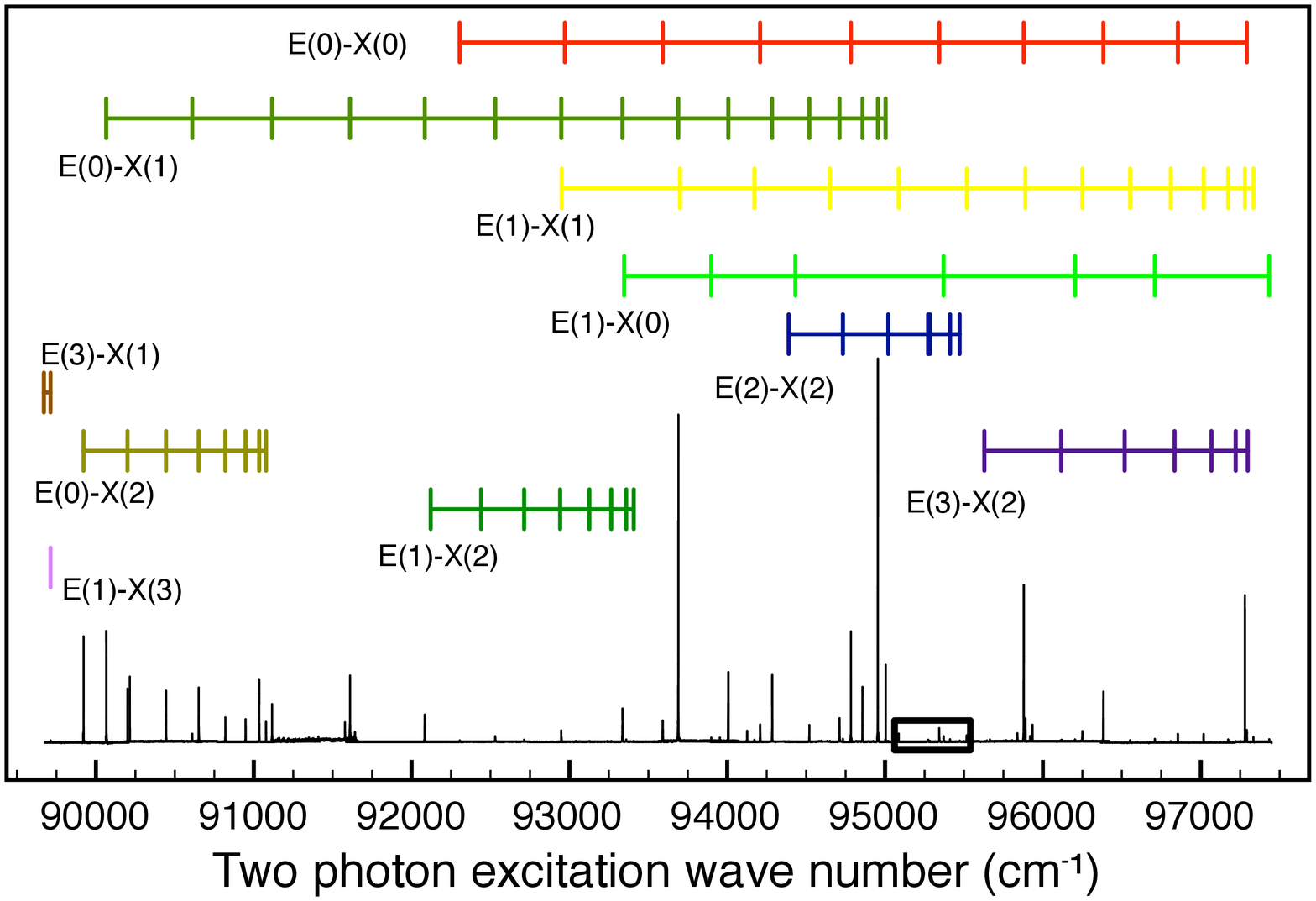}
\caption{(Colour online) Doppler-free (2+1) REMPI spectrum of the H$_{2}$ $EF^{1}\Sigma_g^+$-$X^{1}\Sigma^{+}_{g}(v',v")$ bands between 90\,000 and 97\,000 \wn. All assigned transitions belong to the $Q$-branch of their respective band. The small rectangle indicates the range shown in Fig.~\ref{Fig:AssignmentsDetail}.}
\label{Fig:Assignments}
\end{figure}

\begin{figure}[t]
\includegraphics[width=1\linewidth]{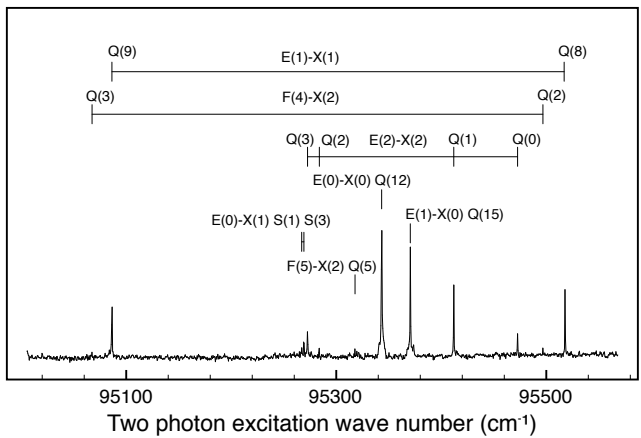}
\caption{Detail of Fig.~\ref{Fig:Assignments} showing several transitions to both the inner, $E$, and outer, $F$, wells. $S$ branch transitions ($\Delta J=2$) were observed for bands with favourable Franck-Condon overlap, in this case the $E(0)$-$X(1)$ band.}
\label{Fig:AssignmentsDetail}
\end{figure}

Two-photon survey spectra of the $EF-X$ system were recorded in the excitation range 90 000 - 97 000 \wn\ with the PDL system in a Doppler-free geometry with counter-propagating laser beams. Figure~\ref{Fig:Assignments} shows many of the  $> 100$ observed $Q$-branch transitions in the $EF-X$ system. Lines are assigned with the convention followed by Bailly~\etal~\cite{Bailly2009} indicating the $EF(v=0)$, $EF(v=3)$, and $EF(v=6)$ vibrational levels of the inner well as $E(0)$, $E(1)$, and $E(2)$ respectively. The $EF(v=9)$ level is referred to as $E(3)$ following Ref.~\cite{Bailly2009} although this assignment is ambiguous if the wave-function composition is considered (see Sec.~\ref{Sec:MQDT}).
%The vibrational levels of the outer well EF($v=1$), EF($v=2$), EF($v=4$), EF($v=5$), EF($v=7$) and EF($v=8$) are %referred to as F(0), F(1), F(2), F(3), F(4) and F(5) respectively (Have you seen tose, will you discuss those ?)
The weak part of the spectrum  in the rectangle indicated in Fig.~\ref{Fig:Assignments} is enlarged in Fig.~\ref{Fig:AssignmentsDetail}, demonstrating the large dynamic range of signal strengths probed in this investigation.

Figure~\ref{Fig:AssignmentsDetail} shows a narrow range of the spectrum covering some $\sim$500 \wn\ where $Q$ and $S$-branch transitions as well as lines connecting to the $F$ outer well state are observed. $S$-branch transitions are notoriously weak, but could be observed with the PDL system in the (0,1) and (0,2) bands, which have favourable Frank-Condon factors~\cite{FernandezAlonso2000}. However, these $S$-transitions were limited to low, odd $J$ only, as it is enhanced by the ortho-para ratio.

All measured transition frequencies from the survey spectra are accurate to within 0.1 \wn\, and are available with assignments in an electronic document as Table III of the Supplemental Material~\cite{SuppHBrPRA}.
%Included in the document is a list of the unassigned transitions that will be discussed below.
The two-photon transitions from the survey spectra were assigned via three methods: (i) by measuring ground state energy splittings between $O$, $Q$, and $S$ lines in a single band, as well as by ground state vibrational splittings between $Q$-lines; (ii) by making use of the highly accurate $EF^{1}\Sigma^{+}_{g}$ level energies from Bailly~\etal~\cite{Bailly2009} and the ground state level energies~\cite{Komasa2011}; and (iii) by comparing to results from the present MQDT calculations.

The $E(0)$ and $E(1)$ level energies were confirmed experimentally by measuring ground state rovibronic energy splittings, using both the PDL and PDA systems, and comparing to the calculations of Komasa \etal~\cite{Komasa2011}. For low $J$ levels belonging to the $E(2)$, $E(3)$ and $F(4)$ vibrations the combination of the level energies measured by Bailly \etal~\cite{Bailly2009} and the ground state calculations leads to an unambiguous assignment.

A total of 30 transitions remained and were assigned by using the level energies from the MQDT calculation and the ground state calculations. For each of these transitions the ground state rotational energies for $v=0-3$ were added systematically to produce a level energy which was then compared to the MQDT calculation. Levels matching within about 5 \wn\ were considered. Inspection of Table 1 of the Supplemental Material indicates that (with three exceptions labelled "tentative" in the table) the calculated wave function associated with each of these energies contains significant $s$  and/or $d$ Rydberg channel inner-well character, thus making them accessible in vibronic transitions from the ground state. This feature provides further support for the correctness of the assignments. The majority of these transitions belonged to high $J$ states with strong $F$ character occurring near the crossings between $E$ and $F$ rotational levels as is indicated in Fig.~\ref{Fig:LevelEnergiesMQDT}. Furthermore high-$J$ levels belonging to the $E(2)$, $E(3)$, and $EF(10)$ levels were also observed. Finally, of these 30 transitions, four additional transitions correspond to upper-state levels that lie above the potential energy barrier between the inner and outer wells; these are indicated in Table~\ref{Tab:ELevels}.

\section{$EF^{1}\Sigma_{g}^{+}$ Level Energies}
\label{Sec:EnergyLevels}
Tables~\ref{Tab:ELevels} and \ref{Tab:FLevels} present the $EF$-level energies, for the inner and outer wells, respectively, determined from adding the rovibrational excitation energies~\cite{Komasa2011} to the measured transition frequencies. Where possible, resulting level energies (and their uncertainties) are averaged over various measurements of $O$, $Q$ and $S$ branches as well as over lines in vibrational bands. Included in Table~\ref{Tab:ELevels} are some highly accurate levels for $v=0$ and $J \leq 5$ determined from a study with a more narrowband laser source~\cite{Hannemann2006,Salumbides2008}, as well as the accurate level energies determined by Bailly~\etal~\cite{Bailly2009}.
%The contributions to the combined error is dominated by the measurements since the level of accuracy of the %calculations is %estimated at 0.001 \wn . Hence the combined error in the level energies for levels based on the PDL %measurements is 0.1 \wn\ %and 0.005 \wn\ for levels based on the PDA measurements.

When comparing a total of 38 measured $EF^{1}\Sigma_{g}^{+}$ energy levels obtained with the PDL system to those obtained by Bailly~\etal~\cite{Bailly2009}, a systematic deviation of +0.085 \wn\ between the data sets is found, which is attributed to an ac-Stark shift effect. In the metrology measurements with the PDA system it was found that for input energies of between 200 and 400 $\mu$J the ac-Stark shift amounts to 0.005 \wn. The PDL-based survey measurements were performed at an intensity in the focus increased by a factor of 16, which is consistent with an ac-Stark shift of ~0.08 \wn. The level energies presented in Table~\ref{Tab:ELevels} recorded with the PDL system have been corrected for this ac-Stark shift.

We find a difference of 2.95 \wn\ between our value for the $E(1)$ $J=6$ level energy and the value reported in Ref.~\cite{Bailly2009}. The measured level energy derives from the $E-X$ (1,1) $Q(6)$  and $E-X$ (1,0) $Q(6)$  transitions, which in combination give an unambiguous identification. This strongly suggests a mis-assignment for the $E(1)$ $J=6$ level in Ref.~\cite{Bailly2009}, in addition to the $E(0)$ $J=13$ level which was also shown to be a misassignment~\cite{Salumbides2011}. Furthermore the assignment of the $F(3)$ $J$=7 level differs by 14.43 \wn . This transition is based on the $F-X$ (3,1) $Q(7)$ transition which is assigned based on the MQDT calculations. Since the MQDT calculations are accurate to within $\sim$5 \wn\ this is suggestive of a misassignment although experimental verification is needed to confirm this. In the present experiment this is not possible and hence we mark this assignment as tentative. 
\begin{figure}[b]
\includegraphics[width=1\linewidth]{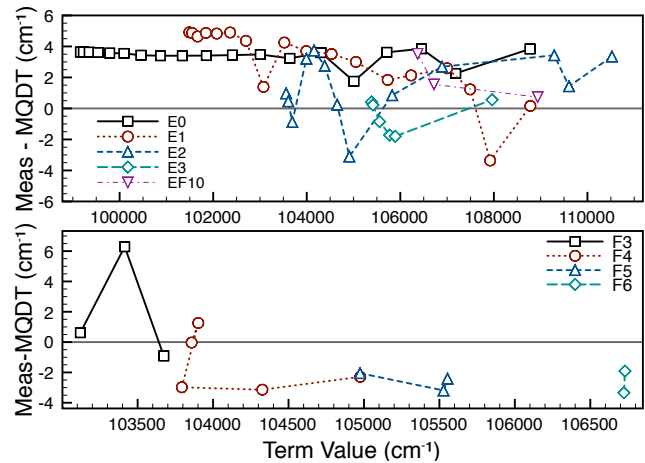}
\caption{(Colour online) Differences between the measurement and the MQDT calculation for the inner-well levels (upper figure) and the outer-well levels (lower figure).}
\label{Fig:MQDTComparison}
\end{figure}

The MQDT calculations of level energies were used for the identification of those levels which were neither present in the study of Bailly~\emph{et al.}~\cite{Bailly2009}, nor observed in a combination difference. Figure~\ref{Fig:MQDTComparison} shows the difference between experimental and MQDT-calculated level energies. Overall these are $\leq$ 5 \wn . It had been found in the previous Ref. \cite{Ross1994} that while the MQDT calculations typically deviated by a few cm$^{-1}$ from the observed level positions, these residuals turned out to be more or less constant along the rotational progression  associated with a given vibronic level (e.g. $E(v')$ or $EF(v'$)), with abrupt changes occurring only near avoided crossings between inner-well and outer-well levels (cf. Table II of Ref. \cite{Ross1994}). As Fig.~\ref{Fig:MQDTComparison} shows, this feature is also borne out rather clearly in the new extended data set listed in the present Table~\ref{Tab:ELevels}. For the $F(4)$ $J=6-8$ levels, comparisons with the current MQDT results and those in Ref.~\cite{Bailly2009} show large discrepancies, in the order of 15~\wn . These levels were not observed in the present study using the PDA or PDL systems but from the evidence presented in Fig.~\ref{Fig:MQDTComparison} this is suggestive of a misassignment, although further experimental evidence would be necessary to confirm this.

\section{Conclusion and Outlook}

A precision metrology study was performed on the $EF^{1}\Sigma^{+}_{g}$-$X^{1}\Sigma^{+}_{g}$ system of H$_2$ via high resolution two-photon Doppler-free spectroscopy. The data were reduced to provide a test of the differential, rotational quantum state dependent, QED and relativistic effects in the $X^{1}\Sigma^{+}_{g}$ ground state of the H$_2$ molecule which were calculated by Komasa~\etal~\cite{Komasa2011}. The agreement is excellent with the theoretical predictions currently more precise than our experiment. In addition survey spectra were recorded in order to obtain an overview of the measurable transitions in the $EF^{1}\Sigma^{+}_{g}$-$X^{1}\Sigma^{+}_{g}$ system, for higher $v$ and $J$ quantum numbers. This has served to extend the known level structure of the $EF^{1}\Sigma^{+}_{g}$ state, thereby paving the way for future studies on QED and relativistic effects in vibrationally excited levels of H$_{2}$. QED calculations show a net decrease for an increase in vibration, with effects of $\sim$0.1 \wn, well within the achievable accuracy of the PDA system. First-principles multi-channel quantum defect calculations were performed on $E$F($v,J$) levels for high angular momentum states up to $J=19$. The calculated level energies were found to deviate less than $\sim$5 \wn\ from the experimentally obtained values and allowed the identification of otherwise unassignable transitions.

\begin{acknowledgements}
This work was supported by the Netherlands foundations for fundamental research on matter (FOM). The authors are thankful to K. Pachucki and J. Komasa for helpful discussions.

\end{acknowledgements}

\bibliographystyle{apsrev4-1}
\bibliography{/Users/garydee/Documents/Articles/CompleteDataBase}

\newpage
\begin{table*}
\caption[]{Table of level energies (in \wn) for the four lowest vibrational levels belonging to the inner well and the first vibrational level above the barrier of the \EFstate\ state derived from the measurements presented here and the ground state levels reported by Komasa~\etal~\cite{Komasa2011}. $\Delta$ represents the difference between the measured levels and the MQDT prediction. Results of Bailly~\etal~\cite{Bailly2009} are included for comparison. See the text for details on the levels grouped under tentative assignments.}
\label{Tab:ELevels}
\begin{tabular}{c r@{.}l c r@{.}l r@{.}l c r@{.}l }
\hline
\multicolumn{1}{c}{$J$} & \multicolumn{5}{c}{$E(0)$ or $EF$($v'=0$)}  & \multicolumn{5}{c}{E(1) or $EF$($v'=3$)}   \\
\hline
\multicolumn{3}{c}{Present Results}&$\Delta$&\multicolumn{2}{c}{Bailly~\emph{et al.}~\cite{Bailly2009}}&\multicolumn{2}{c}{Present Results}&$\Delta$ &\multicolumn{2}{c}{Bailly \emph{et al.}~\cite{Bailly2009}}\\
0	&	99164	&	78691	(11)$^{a}$	&	3.64	&	99164	&	78702(15)	&	101494	&	70	(10)	&	4.91	&	101494	&	74402(15)	\\
1	&	99228	&	21829	(18)$^{a}$	&	3.64	&	99228	&	21824(19)	&	101553	&	95	(10)	&	4.87	&	101554	&	0269(2)	\\
2	&	99354	&	55621	(11)$^{a}$	&	3.63	&	99354	&	55632(14)	&	101671	&	49	(10)	&	4.64	&	101671	&	64197(15)	\\
3	&	99542	&	76607	(15)$^{b}$	&	3.61	&	99542	&	76607(2)	&	101849	&	34	(10)	&	4.87	&	101849	&	4044(2)	\\
4	&	99791	&	32449	(10)$^{b}$	&	3.57	&	99791	&	32519(15)	&	102080	&	91	(10)	&	4.83	&	102081	&	0311(2)	\\
5	&	100098	&	26098	(15)$^{b}$	&	3.55	&	100098	&	26092(2)	&	102367	&	17	(10)	&	4.90	&	102367	&	1451(2)	\\
6	&	100461	&	196	(5)$^{c}$	&	3.43	&	100461	&	19733(2)	&	102704	&	42	(10)	&	4.36	\\
7	&	100877	&	369	(5)$^{c}$	&	3.40	&	100877	&	3708(2)	&	103076	&	23	(10)	&	1.39	\\
8	&	101343	&	824	(5)$^{c}$	&	3.40	&	101343	&	82451(3)	&	103525	&	37	(10)	&	4.25	\\
9	&	101857	&	174	(5)$^{c}$	&	3.41	&	101857	&	17482(2)	&	103994	&	68	(10)	&	3.70	\\
10	&	102414	&	045	(5)$^{c}$	&	3.44	&	102414	&	04588(5)	&	104534	&	09	(10)	&	3.51\\
11	&	103010	&	497	(5)$^{c}$	&	3.48	&	103010	&	49993(3)	&	105052	&	90	(10)	&	3.00	\\
12	&	103641	&	544	(5)$^{c}$	&	3.23	&	103641	&	5428(8)	&	105732	&	04	(10)	&	1.84	\\
13	&	104307	&	453	(5)$^{c}$	&	3.60	&\multicolumn{2}{c}{}&	106232	&	97	(10)	&	2.13	\\
14	&	105009	&	27	(10)	&	1.75	&\multicolumn{2}{c}{}&	107004	&	11	(10)	&	2.59	\\
15	&	105715	&	149	(5)$^{c}$	&	3.62	&\multicolumn{2}{c}{}&	107494	&	19	(10)	&	1.23	\\
16	&	106457	&	320	(5)$^{c}$	&	3.85	&\multicolumn{2}{c}{}&	107917	&	60	(10)	&	-3.36\\
17	&	107186	&	40	(10)	&	2.26	&\multicolumn{2}{c}{}&	108780	&	07	(10)	&	0.16\\																		
19	&	108777	&	75	(10)	&	3.84			\\													
\hline
\multicolumn{1}{c}{$J$}&	\multicolumn{5}{c}{$E(2)$ or $EF$($v'=6$)} & \multicolumn{5}{c}{$E(3)$ or $EF$($v'=9$)}   \\
\hline
\multicolumn{3}{c}{Present Results}&$\Delta$&\multicolumn{2}{c}{Bailly~\emph{et al.}~\cite{Bailly2009}}&\multicolumn{2}{c}{Present Results}&$\Delta$&\multicolumn{2}{c}{Bailly~\emph{et al.}~\cite{Bailly2009}}\\										
0	&	103559	&	58	(10)	&	0.97	&	103559	&	59794(15)	&	105384	&	90	(10)	&	0.42	&	105384	&	9129(2)	\\
1	&	103605	&	61	(10)	&	0.46	&	103605	&	6119(2)	&	105415	&	28	(10)	&	0.23	&	105415	&	2551(2)	\\
2	&	103690	&	18	(10)	&	-0.88	&	103690	&	14695(14)	&\multicolumn{2}{c}{}&		&	105473	&	96704(15)	\\
3	&	103995	&	28	(10)	&	3.21	&	103995	&	2119(2)	&	105556	&	77	(10)	&	-0.85	&	105556	&	8403(2)	\\
4	&	104159	&	81	(10)	&	3.74	&	104159	&	80598(15)	&\multicolumn{2}{c}{}&		&	105657	&	7072(3)	\\
5	&	104386	&	80	(10)	&	2.74	&	104386	&	8711(2)	&	105770	&	05	(10)	&	-1.72	&	105770	&	1314(3)	\\
6	&	104650	&	43	(10)	&	0.24	&\multicolumn{2}{c}{}&	105890	&	04	(10)	&	-1.80\\
7	&	104908	&	69	(10)	&	-3.10\\
9	&	105826	&	53	(10)	&	0.85	\\
11	&	106894	&	85	(10)	&	2.69\\
15	&	109293	&	90	(10)	&	3.43	&\multicolumn{2}{c}{}&	107962	&	14	(10)	&	0.56	\\
16	&	109607	&	40	(10)	&	1.43	\\
17	&	110526	&	86	(10)	&	3.35\\

\hline
\multicolumn{1}{c}{$J$}&	\multicolumn{5}{c}{$EF$($v'=10$)} &\multicolumn{5}{c}{Additional levels above the barrier}\\
\hline
\multicolumn{3}{c}{Present Results}&$\Delta$&\multicolumn{2}{c}{Bailly~\emph{et al.}~\cite{Bailly2009}}&\multicolumn{2}{l}{Present Results}&$\Delta$&\multicolumn{2}{c}{Assignment}\\								 
5	&	106374	&	02	(10)	&	3.52	&106374&1301(3)      &107537	&	79	(10)	&	8.35	& \multicolumn{2}{c}{$EF$(11) $J$=8}	\\
7	&	106721	&	14	(10)	&	1.55	&\multicolumn{2}{c}{} &108386	&	73	(10)	&	1.76	& \multicolumn{2}{c}{$EF$(12) $J$=9}\\
16&	108937	&	93	(10)	&	0.73	&\multicolumn{2}{c}{} &110202	&	77	(10)	&	4.64	&\multicolumn{2}{c}{$EF$(12) $J$=16}	\\
\multicolumn{6}{c}{}                                                           &110806	&	31	(10)	&	4.81& \multicolumn{2}{c}{$EF$(13) $J$=16}	\\
\hline
\hline

\footnotetext[1]{Based on the measurements of Hannemann~\etal~\cite{Hannemann2006}.}
\footnotetext[2]{Based on the measurements of Salumbides~\etal~\cite{Salumbides2008}.}
\footnotetext[3]{Based on the measurements of Salumbides~\etal~\cite{Salumbides2011}.}
\end{tabular}
\end{table*}

\begin{table*}
\caption[]{Table of level energies (in \wn) for all measured levels belonging to the outer well derived from the measured transitions and the ground state calculations reported by Komasa~\etal~\cite{Komasa2011}. $\Delta$ represents the difference between the measured levels and the MQDT prediction. The results of Bailly~\etal~\cite{Bailly2009} are included for comparison.}
\label{Tab:FLevels}
\begin{tabular}{c r@{.}l c r@{.}l c r@{.}l c r@{.}l }
\hline
\multicolumn{1}{c}{$J$} & \multicolumn{5}{c}{$F(3)$ or $EF$($v'=5$)}  &$J$& \multicolumn{5}{c}{$F(4)$ or $EF$($v'=7$)}   \\
\hline
\multicolumn{3}{c}{Present Results}&$\Delta$ &\multicolumn{2}{c}{Bailly~\emph{et al.}~\cite{Bailly2009}}&\multicolumn{3}{c}{Present Results}&$\Delta$&\multicolumn{2}{c}{Bailly~\emph{et al.}~\cite{Bailly2009}}\\
7\footnotemark[20]	&	103121	&	01	(10)	&	0.62	&\multicolumn{2}{c}{} &	1	&	103857	&	92	(10)	&	-0.03	&	103857&	8468(2)	\\
10\footnotemark[20]	&	103420	&	55	(10)	&	6.29	&\multicolumn{2}{c}{}&	2	&	103903	&	07	(10)	&	1.26	&	103902&	9828(3)	\\
12\footnotemark[20]	&	103672	&	48	(10)	&	-0.91	&\multicolumn{2}{c}{}&	3	&	103790	&	09	(10)	&	-2.98	&	103789&	9773(2)	\\
\multicolumn{6}{c}{} &	9	&	104323	&	32	(10)	&	-3.14			\\
\multicolumn{6}{c}{}	 &	14	&	104973	&	02	(10)	&	-2.29		\\
\hline
\multicolumn{1}{c}{$J$}&	\multicolumn{5}{c}{$F(5)$ or $EF$($v'=8$)} &$J$& \multicolumn{5}{c}{$F$(6)}   \\
\hline
\multicolumn{3}{c}{Present Results}&$\Delta$&\multicolumn{2}{c}{Bailly~\emph{et al.}~\cite{Bailly2009}}&\multicolumn{3}{c}{Present Results}&$\Delta$&\multicolumn{2}{c}{Bailly~\emph{et al.}~\cite{Bailly2009}}\\	
5	&	104971	&	99	(10)	&	-2.05	&	104972	&	0087(3)	&	13	&	106727	&	84	(10)	&	-1.91	\\			
11	&	105523	&	80	(10)	&	-3.19	&\multicolumn{2}{c}{}&	14	&	106719	&	61	(10)	&	-3.34	\\			
12	&	105551	&	50	(10)	&	-2.41														\\								
\hline
\hline

\footnotetext[20]{Tentative assignment.}
\end{tabular}
\end{table*}

\end{document}